# ScaleLat: A chemical structure matching algorithm for mapping atomic structure of multi-phase system and high entropy alloys

Nan Li [a], Junming Guo [a], Sateng Li [a], Haoliang Liu [a], Qianwu Li [b], Fangjie Shi [b*], Yefei Li [c*], Bing Xiao [a*]

[a] School of Electrical Engineering, State Key Laboratory of Electrical Insulation and Power Equipment, Xi'an Jiaotong University, Xi'an 710049, China.

[b] Life Management Center, Suzhou Nuclear Power Research Institute, Suzhou 215004, China

[c] State Key Laboratory for Mechanical Behavior of Materials, Xi'an Jiaotong University, Xi'an 710049, China

* Corresponding author, E-mail: shifangjie@cgnpc.com.cn, liyefei@xjtu.edu.cn, bingxiao84@xjtu.edu.cn

**Abstract:** ScaleLat (Scale Lattice) is a computer program written in C for performing the atomic structure analysis of multi-phase system or high entropy alloys (HEAs). The program implements an atomic cluster extraction algorithm to obtain all independent and symmetry-reduced characteristic chemical structures for the complex atomic configurations which are usually obtained from molecular dynamics or kinetic Monte-Carlo simulations for supercell containing more than $10^4$ atoms. ScaleLat employs an efficient and unique chemical structure matching algorithm to map all extracted atomic clusters from a large supercell (>$10^4$ atoms) to a representative small one (~ $10^3$ or less), providing the possibility to directly use the highly accurate quantum mechanical methods to study the electronic, magnetic, and mechanical properties of multi-component alloys with complex microstructures. We demonstrate the capability of ScaleLat code by conducting both the atomic structure analysis and chemical structure matching procedure for Fe-12.8 at.% Cr binary alloy and equiatomic CrFeCoNiCu high entropy alloy, and by successfully obtaining the representatively supercells containing $10^2$~$10^3$ atoms of the two alloys. Overall, ScaleLat program provides a universal platform to efficiently project all essential chemical structures of large complex atomic structures to a relatively easy-handling small supercell for quantum mechanical calculations of various user interested properties.

**Program summary**

*Program* Title: ScaleLat

*CPC Library* link to program files:

*Developer's* repository link: https://github.com/NanLi-xjtu/ScaleLat.git

*Licensing* provisions: MIT

*Programming* language: C

*Nature of Problem:* Very large supercells containing more than $10^4$ atoms must be used to describe the atomic structure of complex multi-phase or high entropy alloys, and their properties are almost not possible to investigate using quantum mechanical calculations within the conventional computational facilities right now. Decreasing the total number of atoms in a smaller representing supercell which preserves all essential chemical structures of the original large supercell is the key to the problem.

*Solution to the Problem:* An atomic cluster extraction algorithm is realized to thoroughly analysis all essential chemical structures of multi-phase system or high entropy alloys; Utilizing the direct atom swapping method, the chemical structure matching procedure is developed to map all extracted characteristic atomic clusters of benchmark structure to the small supercell by minimizing their differences in both the types and relative proportions of atomic cluster sets.

# 1. Introduction

Metallic alloys consisting of two or more principal components are widely used as both structural and functional materials due their outstanding mechanical, chemical and physical properties [1,2,3,4,5]. For example, the binary ferrite Fe-Cr steels with various Cr contents have been employed to fabricate the pipelines and key components in turbine engines in thermal and nuclear power plants because of their high mechanical toughness, high oxidation resistance and radiation tolerance [6,7,8]. Nowadays, there are also great interests to prepare and explore the properties of multi-component alloys with at least three different constituting metals or even more principal metal components. Alloying multiple principal elements could either result in a multi-phase microstructure or lead to the formation of medium and high entropy alloys (MEA and HEA). The experimental study of high entropy alloys was pioneered by the works of Yeh [9] and Cantor [10] in 2004 on multicomponent equiatomic metallic alloys. Thermodynamically, the formation of MEA or HEA is mainly attributed to the high configurational entropy in the fully disordered multicomponent solid solution structure. So far, experimentally investigated and verified HEAs are highly limited to quite few systems including VNbMoTaW [11,12], TiV$_x$ZrNbMo$_y$ [13], Al$_x$CrFeCoNi [14,15] and CrMnFeCoNi [16,17]. Typically, high entropy alloys could exhibit the good ductility, high mechanical moduli (Young's modulus and shearing modulus), high yield strength and fracture toughness, and good creep or fatigue resistance. Therefore, HEAs are very often considered as the promising structural materials. Nevertheless, employing HEAs as the novel catalysts for hydrogen evolution reaction (HER), oxygen evolution reaction (OER) and also magnetic materials have also been reported [18,19].

Investigating the atomic structures of multi-component alloys is essential to reveal the complexity in their interatomic chemical interactions, and which is also critical to establish possible quantitative relationships between their properties and compositions or microstructure. Computer simulations based on first-principles calculations or classic molecular dynamics simulations play the vital role together with experimental study in understanding those aspects. The first-principles calculations, which are usually based on density functional theory, are considered as non-empirical method. Applying the first-principles method directly to multi-component alloys with a multi-phase microstructure or fully disordered lattice structure faces great challenges, because both the time and spatial scales handled in typical quantum mechanical calculation are highly limited due the high computational costs and complexity in algorithm [20,21,1]. For example, to properly simulate the atomic structure evolution in Fe-Cr binary alloy at the typical phase separation composition (> 10.0 at.% Cr) and annealing temperature (600-800 K), it requires a supercell model containing at least $10^5$ atoms [22,23,24]. Otherwise, addressing the microstructure of HEAs has been done on a simulation cell containing more than $10^4$ atoms [1]. Therefore, the less accurate interatomic potentials are generated using the training and validation datasets obtained from first-principles calculations on structural, stress and energy parameters. The interatomic potentials are then combined with either classic molecular dynamics simulations or Monte-Carlo method to initiate the atomic structural evolution in multi-component alloys within $10^6$ atoms on any modern-day workstation, providing a highly efficient theoretical methodology to qualitatively study the interatomic bonding mechanism and the microstructure at atomic scale. However, reliability and accuracy of interatomic potentials are critically dependent on the quality of training and validation datasets. In addition, the spatial scale that is usually produced using classic molecular dynamics simulation or Monte-Carlo simulation is many orders of magnitude higher than the typical supercell size which could be efficiently handled by quantum mechanical calculations. As a result, adopting first-principles method to predict the elastic constants, mechanical moduli, electronic density of states and magnetic ordering for multi-component alloys is still considered as a formidable task nowadays.

In this paper, we proposed a novel chemical structural mapping and matching methodology to efficiently and accurately describe the atomic structure of multi-component or multi-phase alloys by directly projecting the microstructure obtained from the molecular dynamics simulations or Monte-Carlo method in long time and large spatial scales to small supercell model within less than 1000 atoms. Notably, for such a small supercell, the first-principles methods may be used to calculate the mechanical, physical, and chemical properties. Our methodology partly resembles that of cluster expansion in terms of characteristic atomic cluster extraction algorithm. However, the proposed method deviates from that of cluster expansion in the sense that the additional atomic structure matching algorithm is also integrated with the atomic cluster mapping process, enabling the user to physically rescale the dimensions of the initial large supercell to a specified size containing only few hundred atoms. Besides the aforementioned rescaling algorithm for supercell dimensions, the algorithm also provides additional atomic structural analysis abilities including the Warren-Cowley short-range order parameters and atomic structure files for all symmetry-reduced chemical clusters. Here, we implemented the algorithm in a self-contained software package named ScaleLat. ScaleLat is considered as a very useful tool to further boost our ability to use quantum mechanical methods for investigating the physical and chemical properties of HEAs and multi-phase structures in different fields.

The paper is organized as follows. In Section 2, the workflow of ScaleLat and the implementation of structural mapping and matching algorithm are discussed. The use of ScaleLat is illustrated in Section 3, employing Fe-Cr binary alloys and CrFeCoNiCu alloy as examples. The conclusion is presented in Section 4.

## 2. Methodology

### 2.1 ScaleLat workflow

The workflow of ScaleLat software is illustrated in Figure 1. The implementations of all relevant methodology and algorithm are based on C. Therefore, the source code of ScaleLat can be easily complied and installed on any operation system with a proper C compiler, i.e., C/GCC on Linux or Windows system. The current version of the software can efficiently realize the extraction and mapping the characteristic and symmetry irreducible atomic clusters for supercell models of multi-component alloys containing $10^4 \sim 10^6$ atoms using the laptop or desktop computer. The atomic cluster extraction algorithm currently implemented in ScaleLat code outputs all symmetrically non-equivalent atomic clusters in terms of cluster sets. The cluster sets are further assigned to different categories by the order of the atomic concentration of the principal constituting elements. Clearly, those cluster sets represent the characteristic chemical structure units of the large supercell for multi-component alloys. The cluster matching algorithm employs all cluster sets as both the benchmarks and the training datasets to optimize the atomic configuration of the user defined small supercell. Eventually, ScaleLat enables a direct projection of all characteristic chemical structures of a big supercell to a small supercell that could be handled with more accurate first-principles methods.

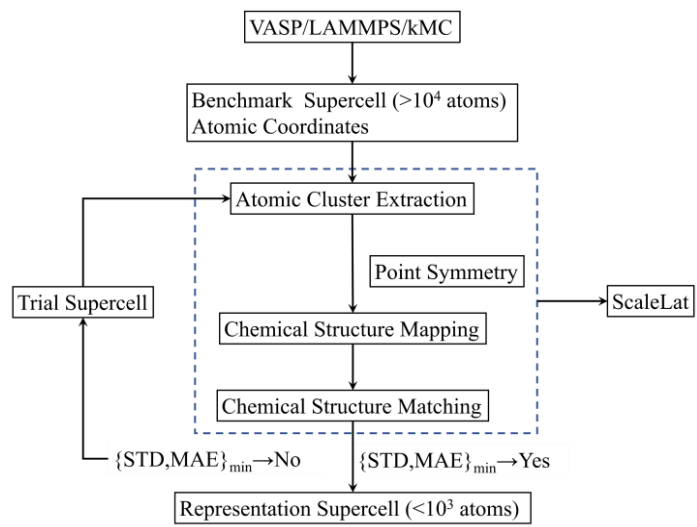

**Figure. 1.** The workflow of ScaleLat software

ScaleLat uses the supercell models of multi-component alloys obtained from other theoretical or experimental methods as the only input data for the subsequent atomic cluster extraction and chemical structure matching calculation. The standard input file requires the atomic coordinates and lattice vectors of supercell for multi-component alloys. Therefore, ScaleLat can be easily interfaced to other commercial or open-source computational tools for atomic structure modelling based either on classic molecular dynamics simulations or first-principles calculations, including LAMMPS, CASTEP, VASP, Quantum Expresso and etc. The atomic structure obtained from kinetic Monte-Carlo simulation is also the scope of ScaleLat software. The users can easily use their own scripts or computer codes to directly convert other atomic structure file formats into the one which could be processed by ScaleLat code.

**2.2 Implementation**

The functionality of ScaleLat software relies on two major computational modules including cluster extraction module and cluster mapping module, as shown in Figure 2.

Cluster extraction module picks up an atomic cluster centered at a randomly selected lattice site in the supercell model. The size of the atomic cluster is defined by a cutoff radius ($r_{cut}$). Users are allowed to use different $r_{cut}$ values for the task. It is worth mentioning that the computational costs are expected to increase rapidly with the increase of $r_{cut}$ value. For each extracted atomic cluster, the atomic percentages of all constituting elements are evaluated. A principal element is assigned to the atomic cluster, allowing each atomic cluster to be allocated to a specific atomic cluster set. For any multi-component atomic cluster, it could be assigned to different atomic cluster sets simultaneously, depending on possible options for principal element. Then, the point symmetry operations are employed to atomic registries of all clusters, removing all symmetrically equivalent atomic clusters in the atomic cluster set. The symmetry reduction of the cluster set is realized using Eq. (1), where X and R represent the set of atomic coordinates and the point symmetry operations, respectively.

$$\left.\begin{aligned} R \otimes X \{r_1, r_2, \ldots \ldots r_N\} = X \{R \otimes r_1, R \otimes r_2, \ldots \ldots R \otimes r_N\} = X' \\ X = X', \text{ symmetry equivalent} \\ X \neq X', \text{ symmetry non-equivalent} \end{aligned}\right\} \quad (1)$$

Regarding the symmetry reduction step, the current algorithm only considers the point symmetry operations of simple cubic and orthorhombic point groups. Specifically, for simple cubic symmetry, the point group includes $C_2$ and $C_4$ rotational axes in x, y, and z directions. Meanwhile, the simple orthorhombic point group contains $C_2$ axes in x, y, and z directions. Other symmetry elements such as mirror planes, inversion center and rotation-inversion operation are not considered, because those point operations are not expected to be seen very often in HEAs or multi-phase microstructures. Otherwise, the employed point operations in the current algorithm are sufficient to perform the symmetry analysis for all extracted atomic clusters in multi-component alloys. In the final step, the cluster extraction module outputs all symmetrically non-equivalent atomic clusters obtained from the input supercell model by the order of the atomic concentration of principal constituting elements. For a typical multi-component alloy, the atomic cluster set may usually consist of $10^2 \sim 10^3$ cluster configurations (chemical structure types).

The cluster mapping module performs a chemical structure matching algorithm by mapping the symmetry reduced cluster set to a user defined small supercell usually containing no more than $10^3$ atoms. This trial supercell is defined with respect to a unit cell by $N_1 \times N_2 \times N_3$, where $N_i$ represents an integer. The starting chemical composition of the smaller trail supercell is built as close to that of benchmark supercell as possible.

The chemical structure matching algorithm realized in the ScaleLat software ensures that all symmetry independent atomic clusters obtained from large supercell model are properly mapped onto the user defined small supercell. This goal is achieved by implementing a self-consistent chemical structure matching algorithm that uses a direct atomic swapping technique to update the atomic configuration of small supercell in each iteration according to the benchmark atomic cluster set. During each structural iteration, the atomic cluster set is obtained for the updated atomic configuration of small supercell using cluster extraction module. Then, the standard deviation (STD) and mean absolute error (MAE) are calculated, as given by Eq. (2).

$$\left.\begin{aligned} STD &= \frac{1}{N}\sum_{i=1}^{N}\left(p_i - p_i^0\right)^2 \\ MAE &= \frac{1}{N}\sum_{i=1}^{N}\left|p_i - p_i^0\right| \end{aligned}\right\} \quad (2)$$

A successful atomic structure update is implied by the reduction of both STD and MAE for small supercells. Otherwise, the updated atomic configuration is rejected in the iteration, and the direct swapping of a randomly selected atomic pair in the structure is repeated. The iteration process terminates when the changes of STD and MAE are smaller than some specified numerical criteria, i.e., $(STD)_{n+1} - (STD)_n < \varepsilon$ and $(MAE)_{n+1} - (MAE)_n < \varepsilon$, and n denotes the total number of iterations. Obviously, one should not expect the obtained STD and MAE values of a single trial calculation would meet the global convergence criteria. Therefore, several different smaller supercells are created automatically in cluster matching step, evaluating STD and MAE values for each trial supercell by repeating the above structural updating and cluster mapping procedures. Once the STD and MAE are reduced to the user defined global threshold values, the convergence is reached for the chemical structure matching algorithm, and the calculation is terminated. ScaleLat outputs the STD and MAE values for all sample small supercells, and the atomic structures of them at the last iteration for atomic swapping step are also provided in CONTCAR format.

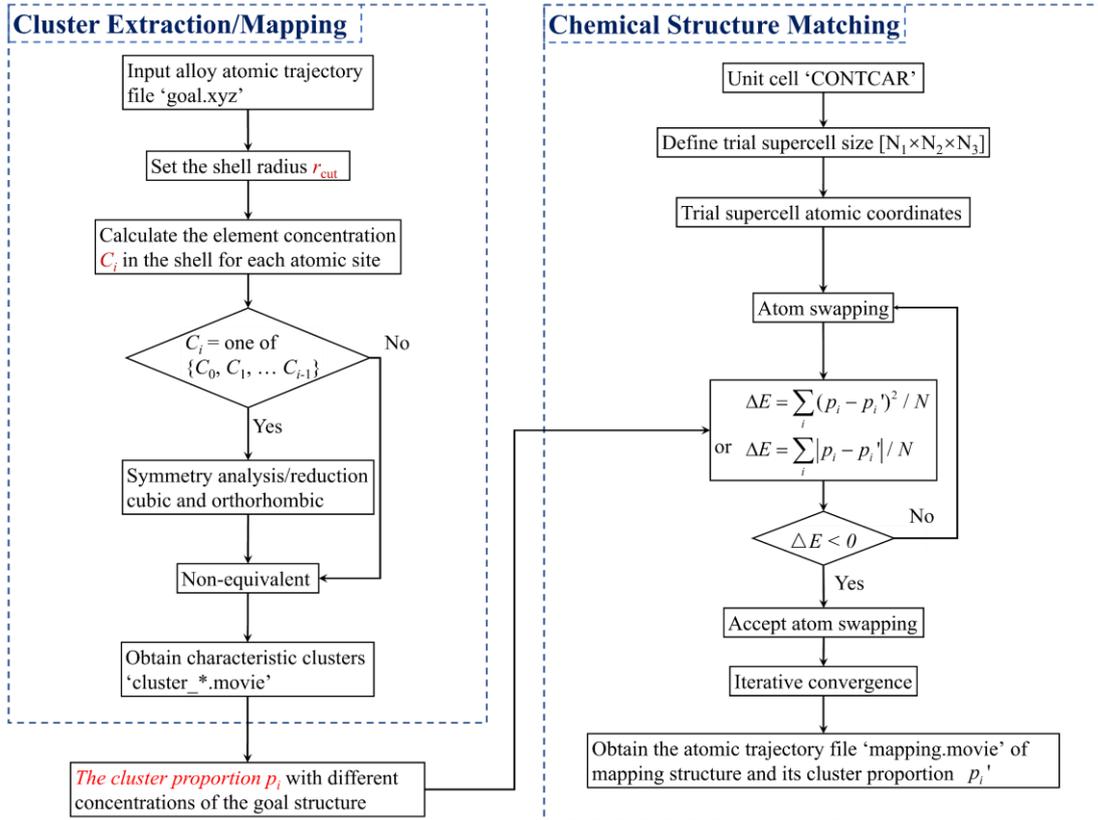

**Figure. 2.** Implementation details of ScaleLat software for mapping and matching chemical structures in multi-component alloys.

Finally, it is also worth noting that the atomic configuration updating algorithm employed here for the chemical structure matching step in ScaleLat code partly resembles the well-known kinetic Monte-Carlo simulation based on Metropolis method in the sense that the new atomic structure is generated by directly swapping atomic pair in the previous iteration [22,23]. However, ScaleLat does not calculate the total energies of the atomic configurations before and after the exchanging of atomic positions. The acceptance of the new atomic configuration in ScaleLat code is simply determined by a successful reduction of STD and MAE values, compared to those of previous step.

## 3. Examples

Here, we would like to demonstrate both the usage and capability of ScaleLat code for analyzing the chemical structures of multi-phase system and high entropy alloys. For the multi-phase system, the binary Fe-12.8 at.% Cr alloy is used, because it shows a interestingly two-phase separation mechanism between Fe-rich ferrite ($\alpha$) and Cr-rich precipitates ($\alpha'$) by annealing [22,25,26]. At different aging stages, the microstructure of Fe-Cr binary alloy changes significantly from a single solid solution phase to the two-phase system, indicating the strong dependence of characteristic atomic clusters on the aging time. In the case of high-entropy alloy, we choose CrFeNiCoCu with equiatomic concentration for all constituting elements [1,27]. The CrFeNiCoCu alloy exhibits the unique atomic structure ordering of Cr atoms on specific crystallographic planes, leading to the formation of chemically ordered superlattice structure. Notably, for both Fe-12.8 at.% Cr binary alloy and CrFeNiCoCu HEA, the very large supercell within at least $2\times10^3$ atoms is needed to reliably capture the atomic structure evolution in either molecular dynamics simulation or kinetic Monte-Carlo simulation [1]. Therefore, applying first-principles methods to investigate the lattice structures, electronic density of states and mechanical properties of the resulting supercells for Fe-Cr binary alloy or CrFeNiCoCu HEA remains a formidable task even on the modern high-performance computing facility.

**3.1 Fe-12.8 at.% Cr binary alloy**

The Cr content in Fe-12.8 at.% Cr binary alloy is slightly above that of the solubility limit of Cr in ferrite-Fe matrix. Here, we use an in-house code to perform the kinetic Monte-Carlo simulations for Fe-12.8 at.% Cr alloy at 673 K by applying the direct atom swapping mechanism and embedded atom method (EAM) potentials for Cr and Fe atoms [28,24]. The employed supercell has 128000 atoms, and total duration of kMC simulation is about $2.0 \times 10^{10}$ steps. Based on previous theoretical and experimental results, the formation of Cr-rich precipitates in Fe-12.8 at.% Cr alloy is suggested to follow a nucleation-growth (NG) mechanism [23,28]. In Figure. 3, the atomic structures of Fe-12.8 at.% Cr binary alloy at different aging stages are illustrated using a supercell containing 128000 atoms in kMC simulations. The kMC simulation results clearly indicate the gradual growth of Cr-rich precipitates in ferrite-Fe matrix with the increasing of total kMC simulation steps, as shown in Figures 3(a)-3(d). To extract the characteristic atomic cluster from the benchmark large supercell (128000 atoms) using the ScaleLat code, the cutoff radius of the atomic cluster is given as the distance between the central atom and the second nearest neighbors in body centered cubic (BCC) lattice structure, resulting in 15 atoms (6 nearest and 8 second neighbors around each central atom) for each atomic cluster. For Fe-Cr binary alloy, the cutoff radius adopted here is sufficiently large to capture most important interatomic interactions among Fe-Fe, Fe-Cr and Cr-Cr that are relevant to $\alpha$-$\alpha'$ phase separation process. As an example, the chemical structure matching algorithm produces four small supercells which contain 1024 atoms at different aging stages by minimizing STD and MAE values of atomic cluster sets extracted from small supercell with respect to those of benchmark large supercell (128000 atoms). The atomic structures of small supercells of Fe-12.8 at.% Cr alloy are displayed in Figures. 3(e)-1(h). The small supercells also exhibit the clustering of Cr atoms at various stages in kMC simulations, and closely resemble those of large supercells. Interestingly, the mixing of Fe and Cr atoms in the Cr-rich region is also seen in the small supercell at the early stage of aging in kMC simulation (See Figure 3(e)). Therefore, the ScaleLat code correctly reproduces the atomic configuration of Fe-Cr alloy at the early stage of aging, because the fully disordered solid solution is assumed for Fe and Cr atoms in the ferrite matrix as the starting structure.

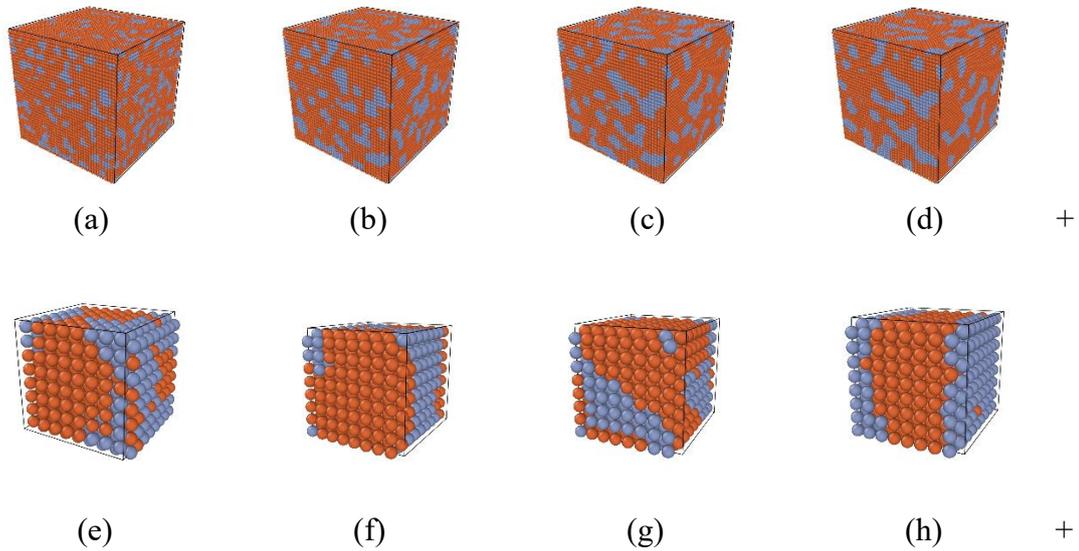

**Figure. 3.** The atomic structures of Fe-12.8 at.% Cr binary alloys at different aging stages in kMC simulations at 673 K using 128000 atoms ((a)-(d)), and the corresponding small supercells containing 1024 atoms obtained from chemical structure matching algorithm in ScaleLat code ((e)-(h)): (a) and (e): $1.0 \times 10^9$ kMC steps; (b) and (f): $5.0 \times 10^9$ MC steps; (c) and (g):

$1.0 \times 10^{10}$ MC steps; (e) and (h): $2.0 \times 10^{10}$ MC steps.

Besides testing the supercell within 1024 atoms, we show the obtained STD and MAE values by varying the supercell sizes from 128 atoms up to 1024 atoms in Figure. 4. The chemical structure matching algorithm does not perform very well for the smallest supercell containing only 128 atoms, leading to the large STD and MAE values. By increasing the total number of atoms in the supercell, the obtained STD and MAE reduce very rapidly with the supercell size. From Figure. 4, we may conclude that the smallest supercell, which could be used to reliably represent the chemical structure of large supercell (128000 atoms), should contain no less than 256 atoms. Nevertheless, both STD and MAE are decreased by increasing the supercell size further, but overall improvement in the accuracy is minor for supercells within 512 and 1024 atoms, compared to that of 256 atoms.

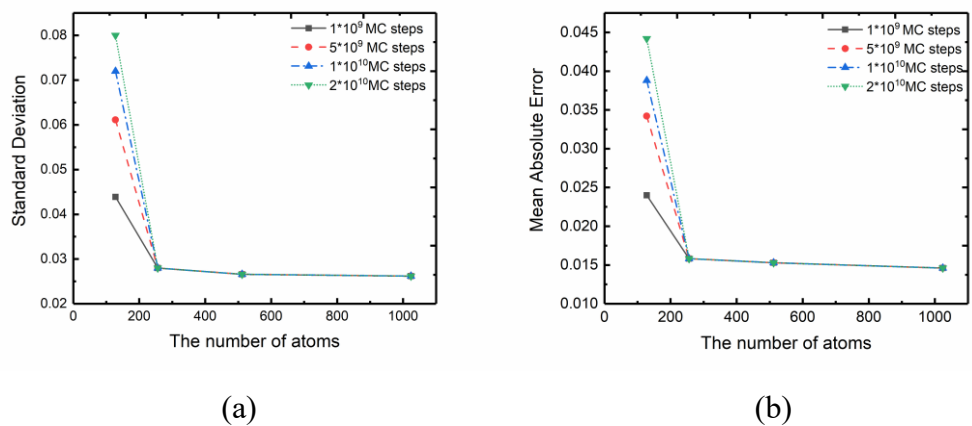

(a)  (b)

**Figure. 4.** Testing the convergence of standard deviation (STD) and mean absolute error (MAE) with respect to the supercell size at different aging stages for Fe-12.8 at.% Cr binary alloy: (a): STDs for 128, 256, 512 and 1024 atoms; (b): MAEs for 128, 256, 512 and 1024 atoms.

To further understand the possible origin of large STD and MAE values when using the smaller supercells in chemical structure matching algorithm, the calculated relative percentages for different atomic cluster sets are depicted in Figure 5. For the smallest supercell, as can be seen from Figure. 5(a), the large deviations are seen between the target supercell (128 atoms) and the benchmark supercell (128000 atoms) for most characteristic atomic clusters in a wide range of Cr atomic concentrations. Notably, the chemical structure matching errors are highest for Fe ferrite matrix and Cr-rich phase among all extract atomic cluster sets. By increasing the total number of atoms in the target supercells, the better cluster matching results are obtained in Figures. 5(b) and 5(c) for supercells consisting of 256 and 512 atoms, respectively. The chemical structure matching quality is significantly improved for Fe-ferrite and Cr-rich phase. In the case of the largest trial supercell (1024 atoms) tested in this work, the overall chemical structure matching quality for all atomic cluster sets is somehow very similar to that of smaller supercell within 512 atoms. Nevertheless, using the larger supercell, the atomic cluster matching errors are indeed decreased further for Cr-rich side (See Figures. 5(c) and 5(d)). If one is particularly interested in using first-principles method to investigate the electronic, magnetic and elastic properties of Fe-12.8 at.% Cr binary alloys at the late stage of aging, a supercell containing 512 atoms is an appropriate choice.

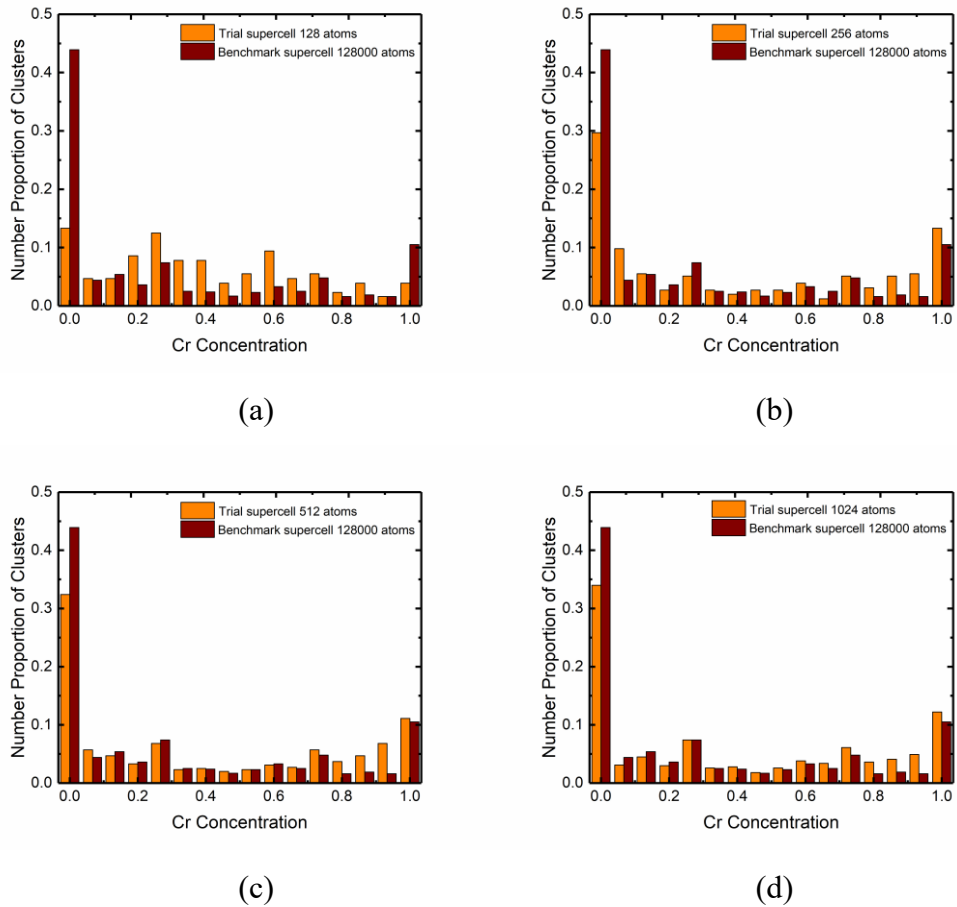

**Figure. 5.** The obtained relative percentages of different cluster sets for supercells containing 128 atoms (a), 256 atoms (b), 512 atoms (c) and 1024 atoms (d) of Fe-12.8 at.% Cr binary alloy

in chemical structure matching algorithm. All atomic clusters having the same atomic concentration for the constituting element Cr are assigned to a single cluster set. Different atomic cluster sets are ordered in the way that the atomic concentration of Cr element is gradually increased from 0.0 to 1.0. The benchmark datasets are obtained from the large supercell (128000 atoms) by running kMC simulation within $2.0 \times 10^{10}$ steps at 673 K.

## 3.2 CrFeCoNiCu high entropy alloy

To generate the benchmark atomic structure of an equiatomic CrFeCoNiCu high-entropy alloy, we first build a large supercell consisting of 108000 atoms with a face-centered cubic like lattice structure. Then, the kinetic Monte-Carlo simulation is carried out at 200 K using EAM potentials for all relevant elements [1,27]. The total duration of kMC simulation is set to $2 \times 10^7$ steps. The atomic structure evolution of CrFeCoNiCu HEA proceeds by employing the direct atom swapping mechanism and the Metropolis algorithm. The obtained atomic structure of CrFeCoNiCu HEA is illustrated in Figure. 6(a), and which is used as the benchmark supercell for chemical structure matching procedure. In Figure. 6(b), we show the fingerprint of all representative chemical structures for CrFeCoNiCu HEA after performing the atomic cluster extraction procedure. Previous theoretical studies on the atomic structures of CrFeCoNiCu HEA or CrCoNi medium entropy alloy (MEA) revealed the possible chemical ordering of Cr atoms in the lattice structure [1,27]. Specifically, Cr atoms tend to segregate on a particular crystallographic plane, forming the superlattice atomic arrangement. In our kMC simulation for CrFeCoNiCu HEA, we find that the obtained microstructure is not uniform, probably due to segregation of multiple constituting elements (Cr, Co, Fe and Cu) at the lower part of the benchmark supercell (See Figure. 6(a)) in kMC simulation at 200 K. For mapping all characteristic chemical structures to the small supercell, the total number of atoms in the supercells is increased in the order of 256 atoms, 480 atoms, 960 atoms and 1920 atoms. In Figure. 6(c), the variations of STDs and MAEs with the different supercell sizes are displayed. Similar to that of Fe-12.8 at.% Cr binary alloy, both STD and MAE are reduced by gradually increasing the total number of atoms in the trial supercells. The calculated STD and MAE have the largest values for the smallest supercell within 256 atoms. By minimizing STDs and MAEs for small supercells in the chemical structure matching step, the obtained atomic configurations are illustrated in Figures. 6(d)-6(f). Clearly, in the case of 256 atoms, the trail supercell is too small to represent all characteristic chemical structures in benchmark large supercell for CrFeCoNiCu HEA. Specifically, the chemical ordering of Cr, Co, Fe and Cu atoms in not well-reproduced in the atomic configuration when using 256 atoms in small supercell. On the other hand, employing the supercells containing 480, 960, and 1920 atoms can reproduce the chemical ordering of the above elements in the benchmark supercell obtained from kMC simulation.

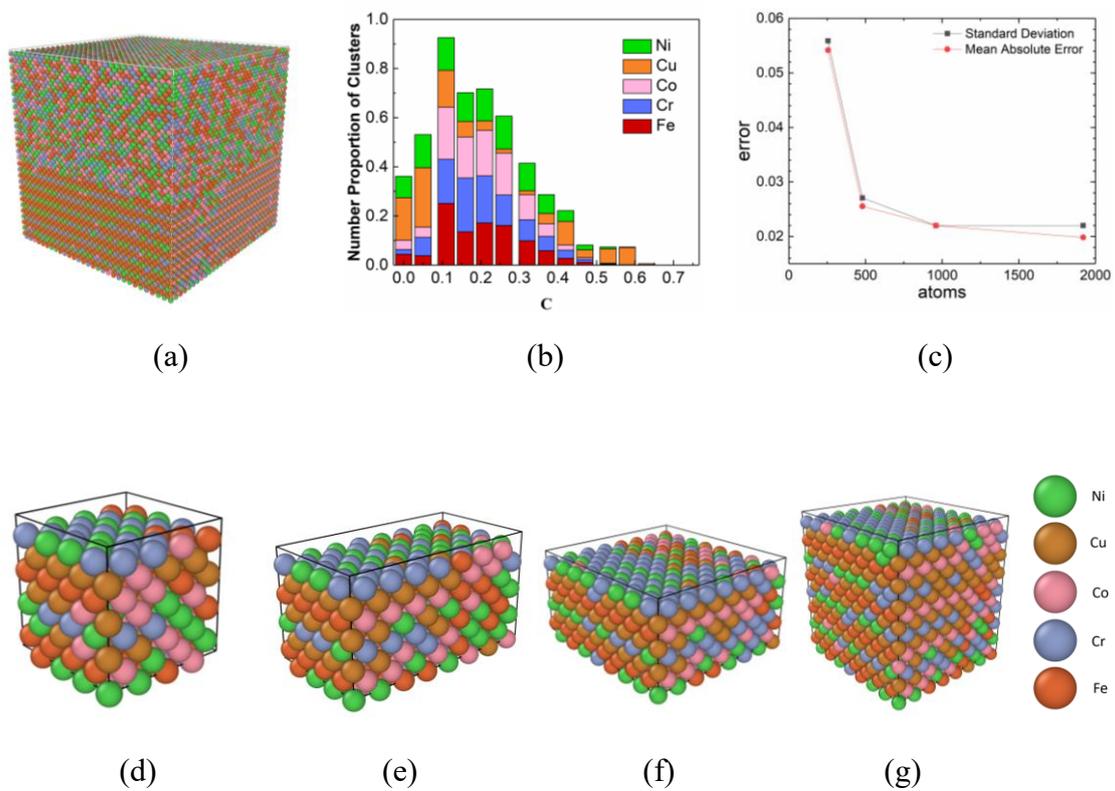

(a)            (b)            (c)

(d)     (e)     (f)     (g)

**Figure. 6**. The atomic structures and characteristic chemical clusters of equiatomic CrFeCoNiCu high entropy alloy: (a): The benchmark large supercell containing 108000 atoms obtained from kinetic Monte-Carlo simulation at 200 K; (b): All characteristic atomic cluster sets are shown by the order of the atomic concentrations of constituting elements; (c): The calculated STDs and MAEs for all small trail supercells; (d)-(f): Atomic configurations of small supercells containing 256, 480, 960, 1920 atoms using chemical structure matching algorithm.

To further evaluate the matching quality of the small supercell for different atomic cluster sets, we compare the relative proportions of the extracted chemical structures for each element in small supercell (1920 atoms) to those of benchmark large supercell

(108000 atoms), and the results are plotted in Figure. 7. Besides Cu, we see a good agreement between the trial small supercell and the benchmark large supercell for all extracted chemical structures in CrFeCoNiCu HEA. Otherwise, even using the supercell containing 1920 atoms could not fully capture all atomic cluster fingerprints of benchmark supercell for Cu-rich chemical structures in CrFeCoNiCu HEA. Nevertheless, most chemical structures of benchmark supercell are indeed reproduced using the small trial supercells containing less than 1000 atoms for CrFeCoNiCu HEA. Beginning with a very large supercell, the ScaleLat code provides a reliable and efficient tool to fold all essential chemical structures of high entropy alloys into an easy handling small supercell by quantum mechanical methods.

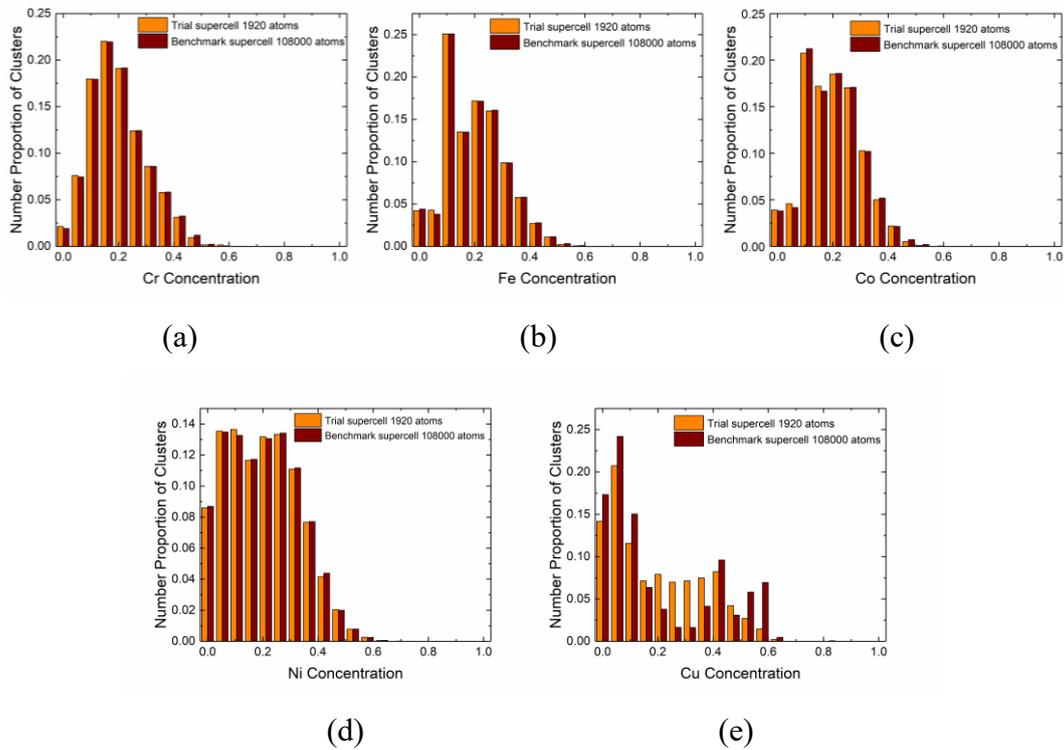

**Figure. 7**. Comparisons of the extracted atomic clusters for different principal elements in small supercell with those of benchmark structure in the case of CrFeCoNiCu HEA: (a)-(e): Cr, Fe, Co, Ni and Cu, respectively. The trail supercell has 1920 atoms.

## 4. Conclusion

ScaleLat is a useful computational tool and universal platform to conduct the chemical structure analysis for multi-phase alloys or multi-component solid solutions based on the fast, efficient, and user-friendly atomic cluster extraction procedure and chemical structure matching algorithms. The most unique feature of ScaleLat is that it allows the user to mapping all characteristic chemical structures of very large supercell containing more than $10^4$ atoms to a small supercell consisting of less than $10^3$ atoms. Employing such a chemical structure matching method, we may use a small supercell to confidently represent the very complex atomic configurations of multi-phase or high entropy alloys obtained usually from molecular dynamics or kinetic Monte-Carlo simulations. More importantly, the reduction of supercell size for complex systems also offers the key advantage that the first-principles methods are appliable to understand the electronic structure, magnetic and mechanical properties for multi-phase structure or high entropy alloys (HEAs).

**Declaration of Competing Interest**

The authors declare that they have no known competing financial interests or personal relationships that could have appeared to influence the work reported in this paper.

**Acknowledgements**

This work is financially supported by the National Key Research and Development Program of China (NO: 2020YFB1901500). Bing Xiao is supported by the "Young Talent Supporting Plan" of Xi'an Jiaotong University (No. DQ1J009). The authors also would like to thank the high performance computing (HPC) center at Xi'an Jiaotong University for supporting the computational resources.

**Appendix**

**Example of Input file for running ScaleLat code.**

Running ScaleLat program requires three input files. Two of them contain the atomic coordinates of the benchmark supercell and the unit cell of trial supercell, respectively. Another remaining input file provides key computational parameters and their values or strings to control the overall accuracy of the numerical results and the contents of output files.

Atomic coordinates of the benchmark supercell are provided by the input file 'filename.xyz'. Using CrFeCoNiCu HEA as an example, the first few lines of 'CrFeCoNiCu_HEA.xyz' are given as follows.

108000    ## atomic number

Lattice="105.600002 0.0 0.0 0.0 105.599999 0.0 0.0 0.0 105.599999"

Properties=species:S:1:pos:R:3    ## box region and the title for ovito

## The following lines are 4 arrays of tabulated values: element_type position_x position_y position_z

Cr 44.0 10.56 12.320001

Cr 82.720001 95.040001 79.199997

Fe 77.440002 33.439999 15.84

Co 22.880001 88.0 29.92

Co 40.48 26.4 59.84

Ni 44.0 0.0 82.720001

... ... ... ...

For all user defined small trail supercells, only the prototype unit cell is required here to execute ScaleLat code. The unit cell is given by the standard 'CONTCAR" format. In the case of CrFeCoNiCu HEA, an FCC unit cell is adopted.

POSCAR file written by OVITO    ## title

1.0    ## scaling factor

    26.3999996185               0.0000000000              0.0000000000

    0.0000000000               28.1599998474             0.0000000000

    0.0000000000              0.0000000000             28.1599998474    ## box region

  Co    Cr    Cu    Fe    Ni    ## elements types

  364   402   435   368   351    ## atomic number

Direct    ## must be fractional coordinates

## The following lines are 3 arrays of tabulated values: position_x position_y position_z

    0.200000003             0.000000000             0.812500000

    0.333333343             0.437500030             0.000000000

    0.866666734             0.125000000             0.562500000

    0.733333349             0.500000000             0.937500000

    0.066666663             0.250000000             0.562500000

0.733333349          0.750000060          0.187500015

… … …

Finally, the job controlling parameters are all provided in the file 'Input', and the contents of 'Input' are attached below using CrFeCoNiCu HEA as the example.

| | | |
|---|---|---|
| cellSize | 1 1 1 | ## The size of cell expansion for small trail supercells |
| Nelems | 5 | ## The number of elements |
| elemName | Co Cr Cu Fe Ni | ## Elements types |
| NLOOP | 1000000 | ## Loop number |
| ith_nebrR | 2 | ## The number of shells extracted by clusters |
| error_map | 1 | ## The standard deviation (1) or the mean absolute error (0) |
| NeighbourMax | 20 | ## The max neighbour number |
| nebrTabFac | 200 | ## the max neighbour number within cutoff radius |

Use command to run the executable file of ScaleLat:

$ ./ScaleLat MAP

**References**

[1] J.P. Du, P. Yu, S. Shinzato, F.S. Meng, Y. Sato, Y. Li, Y. Fan, S. Ogata, Acta Mater. 240 (2022) 118314. https://doi.org/10.1016/j.actamat.2022.118314.

[2] E.P. George, D. Raabe, R.O. Ritchie, Nat. Rev. Mater. 4(8) (2019) 515-534. https://doi.org/10.1038/s41578-019-0121-4

[3] M.H. Tsai, J.W. Yeh, Mater. Res. Lett. 2(3) (2014) 107-123. https://doi.org/10.1080/21663831.2014.912690

[4] M.A. Meyers, K.K. Chawla, Mechanical behavior of materials. Cambridge university press, 2008.

[5] B.S. Murty, J.W. Yeh, S. Ranganathan, P.P. Bhattacharjee, High-entropy alloys. Elsevier , 2019.

[6] G. Bonny, D. Terentyev, L.J.S.M. Malerba, Scr. Mater. 59(11) (2008) 1193-1196. https://doi.org/10.1016/j.scriptamat.2008.08.008

[7] O. Soriano-Vargas, E.O. Avila-Davila, V.M. Lopez-Hirata, N Cayetano-Castro, J.L. Gonzalez-Velazquez, Mater. Sci. Eng. A 527(12) (2010) 2910-2914.

https://doi.org/10.1016/j.msea.2010.01.020

[8] C.A.C. Souza, D.V. Ribeiro, C.S. Kiminami, J Non Cryst Solids 442 (2016) 56-66. https://doi.org/10.1016/j.jnoncrysol.2016.04.009

[9] J.W. Yeh, S.K. Chen, S.J. Lin, J.Y. Gan, T.S. Chin, T.T. Shun, C.H. Tsau, S.Y. Chang, Adv Eng Mater 6(5) (2004) 299-303. https://doi.org/10.1002/adem.200300567

[10] B. Cantor, I.T.H. Chang, P. Knight, A.J.B. Vincent, Mater. Sci. Eng. A 375 (2004) 213-218. https://doi.org/10.1016/j.msea.2003.10.257

[11] O.N. Senkov, G.B. Wilks, D.B. Miracle, C.P. Chuang, P.K. Liaw, Intermetallics, 18(9) (2010) 1758-1765. https://doi.org/10.1016/j.intermet.2010.05.014

[12] O.N. Senkov, G.B. Wilks, J.M. Scott, D.B. Miracle, Intermetallics, 19(5) (2011) 698-706. https://doi.org/10.1016/j.intermet.2011.01.004

[13] Y.D. Wu, Y.H. Cai, X.H. Chen, T. Wang, J.J. Si, L. Wang, Y.D. Wang, X.D. Hui, Mater. Des. 83 (2015) 651-660. https://doi.org/10.1016/j.matdes.2015.06.072

[14] W.R. Wang, W.L. Wang, S.C. Wang, Y.C. Tsai, C.H. Lai, J.W. Yeh, Intermetallics, 26 (2012) 44-51. https://doi.org/10.1016/j.intermet.2012.03.005

[15] W.R. Wang, W.L. Wang, J.W. Yeh, J. Alloys Compd. 589 (2014) 143-152. https://doi.org/10.1016/j.jallcom.2013.11.084

[16] J.Y. He, C. Zhu, D.Q. Zhou, W.H. Liu, T.G. Nieh, Z.P. Lu, Intermetallics, 55 (2014) 9-14. https://doi.org/10.1016/j.intermet.2014.06.015

[17] B. Schuh, F. Mendez-Martin, B. Völker, E.P. George, H. Clemens, R. Pippan, A. Hohenwarter, Acta Mater. 96 (2015) 258-268. https://doi.org/10.1016/j.actamat.2015.06.025

[18] L. Sharma, N.K. Katiyar, A. Parui, R. Das, R. Kumar, C.S. Tiwary, A.K. Singh, A. Halder, K. Biswas, Nano Res. (2022) 1-8. https://doi.org/10.1007/s12274-021-3802-4

[19] G. Zhang, K. Ming, J. Kang, Q. Huang, Z. Zhang, X. Zheng, X. Bi, Electrochim. Acta 279 (2018) 19-23. https://doi.org/10.1016/j.electacta.2018.05.035

[20] J. Ding, Q. Yu, M. Asta, R.O. Ritchie, Proc. Natl. Acad. Sci. U.S.A. 115(36) (2018)

8919-8924. https://doi.org/10.1073/pnas.1808660115

[21] A. Tamm, A. Aabloo, M. Klintenberg, M. Stocks, A. Caro, Acta Mater. 99 (2015) 307-312. https://doi.org/10.1016/j.actamat.2022.117621

[22] P. Erhart, A. Caro, M.S. de Caro, B. Sadigh, Phys. Rev. B 77(13) (2008)134206. https://doi.org/10.1103/PhysRevB.77.134206

[23] E. Martínez, O. Senninger, C.C. Fu, F. Soisson, Phys. Rev. B 86(22) (2012) 224109. https://doi.org/10.1103/PhysRevB.86.224109

[24] G. Bonny, D. Terentyev, L. Malerba, Comput. Mater. Sci. 42(1) (2008) 107-112. https://doi.org/10.1016/j.commatsci.2007.06.017

[25] W. Guo, D.A. Garfinkel, J.D. Tucker, D. Haley, G.A. Young, J.D. Poplawsky, Nanotechnology 27(25) (2016) 254004. https://doi.org/10.1088/0957-4484/27/25/254004

[26] X. Xu, J. Odqvist, M.H. Colliander, M. Thuvander, A. Steuwer, J.E. Westraadt, S. King, P. Hedström, Metall Mater Trans A 47 (2016) 5942-5952. https://doi.org/10.1007/s11661-016-3800-4

[27] O.R. Deluigi, R.C. Pasianot, F.J. Valencia, A. Caro, D. Farkas, E.M. Bringa, Acta Mater. 213 (2021) 116951. https://doi.org/10.1016/j.actamat.2021.116951

[28] S.M. Eich, D. Beinke, G. Schmitz, Comput. Mater. Sci. 104 (2015) 185-192. https://doi.org/10.1016/j.commatsci.2015.03.047